\documentclass[3p]{elsarticle}
\usepackage{graphicx}
\usepackage{subfigure}
\usepackage{amsmath}
\usepackage{amssymb}
\usepackage{psfrag}
\usepackage{braket}
\usepackage{bm}
\usepackage{color}

\usepackage[utf8]{inputenc}
\usepackage[T1]{fontenc}

\biboptions{longnamesfirst}

\begin{document}

\title{Nonequilibrium Green function approach \\to photoionization processes in atoms}

\author{D.~Hochstuhl}
\ead{hochstuhl@theo-physik.uni-kiel.de}
\author{K.~Balzer} 
\author{S.~Bauch}
\author{M.~Bonitz}
\address{Institut f\"{u}r Theoretische Physik und Astrophysik, Christian-Albrechts Universit\"{a}t zu Kiel, 24098 Kiel, Germany}

\date{\today}

\begin{abstract}
We present a quantum kinetic approach for the time-resolved description of many-body effects in photoionization processes in atoms. The method is based on the non-equilibrium Green functions formalism and solves the Keldysh/Kadanoff-Baym equations in second Born approximation. An approximation scheme is introduced and discussed, which provides a complete single-particle description of the continuum, while the atom is treated fully correlated.
\end{abstract}

\maketitle

\section{Introduction}
With the development of ultrashort, high-harmonic generated vacuum and extreme ultraviolet (vuv/xuv) laser pulses the path towards time-resolved observation of electronic dynamics in plasmas \cite{kremp_pre99,haberland_pre01}, atoms and condensed matter has been paved, for an overview see e.g. \cite{Bauer2005}. Recent experiments allow for the investigation of electronic motion and relaxation processes on the attosecond (as) time scale \cite{Corkum2007}. Using as pump-probe techniques by combination of an intense femtosecond infrared (ir) and an ultrashort attosecond xuv pulse it became possible to directly probe electronic relaxation processes in multi-electron atoms \cite{Drescher2002} and xuv-induced electron shake-up processes in the time domain by means of time-resolved strong field tunneling measurements \cite{Uiberacker2007}.

Both experimental scenarios demand for a time-resolved theory of photoionization (PI) in a many-body framework. In previous theoretical investigations, powerful tools have been used, such as non-adiabatic tunneling theory and single-active electron approaches by means of solving the time-dependent Schr\"odinger equation \cite{Kazansky2008,krasovski_prl07}. In a recent paper, these processes were analytically studied in detail \cite{madsen2008}. However, all these approaches neglect the electron-electron interaction, which may have  non-negligible effects,  especially in the presence of strong laser fields~\cite{Uiberacker2007}. A promising concept to address these question is multiconfiguration Hartree-Fock, e.g. \cite{caillat-05}. In this work, we develop an alternative time-dependent many-body approach to atomic PI including electronic correlations which is based on non-equilibrium Green functions.

\section{Theory}
We aim at describing atomic systems which are initially in equilibrium, i.e. $\hat H(t) \equiv \hat H_0$ for $t\leq t_0$, and are disturbed by a time-dependent external potential for $t>t_0$. The equilibrium hamiltonian of $\text{N}$ electrons in the atom is given by (we use atomic units) 
\begin{equation}\label{h0}
\hat H_0  \ = \ \sum_{i=1}^{N} \left\{-\frac{\nabla_i^2}{2} + v(\mathbf r_i) \right\} + \frac 12 \sum_{i\neq j}^{N} w(\mathbf r_i - \mathbf r_j) - \mu \hat N \,,
\end{equation}
where the potential of the nucleus, $v(\mathbf r_i)$, as well as the two-particle Coulomb interaction $w(\mathbf r_i - \mathbf r_j)= |\mathbf r_i - \mathbf r_j|^{-1}$ are assumed to be spin-independent. As we will work in the grand-canonical ensemble, the chemical potential contribution [which appears in the density operator] is subtracted for convenience [last term in Eq.~(\ref{h0})].
For times $t>t_0$, the atom is disturbed by a time-dependent external field, and the hamiltonian is modified:
\begin{align}
\hat H(t) \ = \ \hat H_0 \, + \,\sum_{i=1}^{N} v_{\text{ext}}(\mathbf r_i,t) 
= \sum_{i=1}^{N} h(\mathbf r_i,t) + \frac 12 \sum_{i\neq j}^{N} w(\mathbf r_i - \mathbf r_j) - \mu \hat N 
\,.
\end{align}
With the last equation we defined the total single particle hamiltonian, $h = h_0+v_{\text{ext}}$, which will be used below.
In this paper we consider the perturbation by an electromagnetic wave in dipole approximation, i.e. the field is assumed homogeneous on the scale of the atom, $v_{\text{ext}}(\mathbf r,t) \ = \ - e \, \mathbf E(t) \cdot \mathbf r$. The electric field envelope $\mathbf E(t)$ is assumed to have a Gaussian shape
\begin{align}\label{eq:gauss}
\mathbf {\mathcal E}(t) \ = \ \mathbf {\mathcal E}_0 \cos \bigl[\omega (t-t_{\text{mid}}) \bigr] \exp \left[- \frac{(t-t_{\text{mid}})^2}{2\tau^2}\right] \,,
\end{align}
with a pulse duration $\tau$.

\subsection{Contour Green functions}
\begin{figure}\label{Fig_Keldysh_contour}
\centering
 \includegraphics[width=0.35\linewidth]{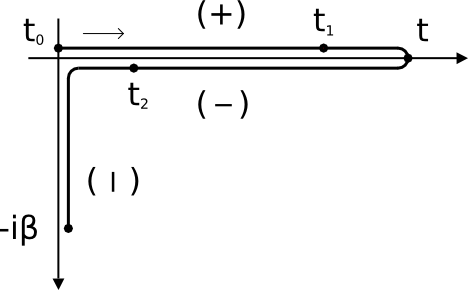}
\caption{The Schwinger/Keldysh contour $\cal C$  running from $t_0$ to $t$, back to $t_0$ and to $t_0-i\beta$ in the complex time plane. $``+'' (-)$ marks the (anti-)chronological real-time branch and $(|)$ the imaginary branch.}
\end{figure}
In the Keldysh/Kadanoff-Baym approach, the central quantity is the one-particle nonequilibrium Green function $G$, which is the time-ordered expectation value of the product of two field operators:
\begin{align}\label{Greenfunction}
G(1,2) \ = \ -i \, \braket{ \, T_{\mathcal C} [ \hat  \Psi_H(1) \hat  \Psi_H^\dagger(2) ] \, } \,,
\end{align}
where the variable $1= (\mathbf r_1,\sigma_1,t_1)$ comprises position, spin projection and time, and the field operators are considered in Heisenberg representation. In the following we denote $\mathbf x_1 =(\mathbf r_1,\sigma_1)$. The ensemble average in (\ref{Greenfunction}) is performed in the grand-canonical ensemble, i.e. with the trace over the unperturbed grand canonical density operator $Z^{-1} e^{-\beta \hat H_0}$. The Green function is defined on the Schwinger/Keldysh time contour ${\cal C}$, see Fig. \ref{Fig_Keldysh_contour}, which allows for an extension of the groundstate and equilibrium formalism and the diagram technique, to non-equilibrium \cite{Keldysh_1964}, for an overview see \cite{Haug_Jauho,bonitz-book}. The contour runs from the initial time $t_0$ to the current time $t$,  back to $t_0$ and, finally to $t_0-i\beta$ in the complex plane where $\beta = 1/k_BT$. The time-ordering operator $T_{\mathcal C}$ in equation (\ref{Greenfunction}) arranges operators with time arguments ``later'' on the contour to the left. The propagation along the complex branch corresponds to the Matsubara formalism, in which the equilibrium density operator is expressed by a time-evolution operator in complex time, $e^{-\beta \hat H_0} \ = \ \hat U(t_0-i\beta,t_0)$. With this, the Green function (\ref{Greenfunction}) takes the explicit form
\begin{align}
G(1,2) \ = \ \frac{\mathrm{Tr}\{\hat U(t_0-i\beta,t_0) \, T_{\mathcal C} [\hat \Psi_H(1) \hat \Psi_H^\dagger(2) ] \, \}}{\mathrm{Tr}\{\hat U(t_0-i\beta,t_0)\} }\,,
\end{align}
in which the time-arguments $t_1$ and $t_2$ each lie on one of the three branches of the contour $\cal C$. It is now convenient to introduce a set of subordinated Green functions, which depend on the location of the two time-arguments on the contour. Accordingly, the Green function becomes a $3 \times 3$ matrix,
\begin{align}\label{gmatrix}
\begin{pmatrix} G^c & G^< & G^\rceil \\ G^> & G^a & G^\rceil \\ G^\lceil & G^\lceil & G^M \end{pmatrix} := \begin{pmatrix} G^{++} & G^{+-} & G^{+|} \\ G^{-+} & G^{--} & G^{-|} \\ G^{|+} & G^{|-} & G^{||} \end{pmatrix}\,,
\end{align}
where $\{+,-,|\}$ mark the position of the respective time-argument on the branches, where the left (right) symbol corresponds to the first (second) time argument. For the notation in the right part of Eq.~(\ref{gmatrix}), see Fig. 1.
The notations in the left part of Eq.~(\ref{gmatrix}) show the relation to the standard definitions, where $G^{\gtrless}$ denote the correlation functions with two real-time arguments and $G^\lceil$ [$G^\rceil$] denotes the correlation function in which the first [second] argument lies on the imaginary branch and the second [first] on one of the real branches. Finally, $G^M$ denotes the Matsubara (imaginary time) Green function of equilibrium theory. The matrix notation allows one to eliminate the time contour $\cal C$ and to consider, in the following, only functions of real time arguments.

Of the four real-time functions $\{G^c,G^<,G^a,G^>\}$, only two are linearly independent, which is why we consider in the following only the correlation functions $G^<$ and $G^>$. The mixed Green functions $G^\lceil$ and $G^\rceil$ account for the evolution of the initial equilibrium state, which itself is determined by the Matsubara Green function $G^M$. As the equilibrium Green function only depends on the difference of two complex time-arguments, we consider the real function $G^M(\mathbf x_1,\mathbf x_2 ,\tau_1-\tau_2) := -i \, G(\mathbf x_1 \, -i\tau_1, \mathbf x_2 \, -i\tau_2)$, which is defined in the interval $[-\beta,\beta]$ and obeys the symmetry $G^M(\mathbf x_1,\mathbf x_2 ,\tau) = -G^M(\mathbf x_1,\mathbf x_2 ,\tau-\beta)$. It is, therefore, sufficient to determine $G^M$ in the range $[-\beta,0]$.

\subsection{Keldysh/Kadanoff-Baym equations}
To compute the time evolution of a multielectron atom, we need to solve the equations of motion of the Green function $G$. The equations for the Keldysh matrix function are the Keldysh/Kadanoff-Baym equations (KKBE) defined on the contour $\cal C$ \cite{Kadanoff_Baym},
\begin{align}\label{KKBE}
\left\{ i\, \partial_{t_1} - h(1) \right\} \, G(1,2) \ = \ \delta_{\mathcal C}(1-2) \, + \, \int_\mathcal{C} \, d3 \; \Sigma[G](1,3) \; G(3,2)\,,
\end{align}
which is to be supplemented by the corresponding adjoint equation. $\Sigma[G]$ denotes the irreducible self-energy which is also a $3 \times 3$ matrix containing mean field (Hartree-Fock) and correlation effects. We will discuss approximations to the self-energy in section \ref{sec:sigma}.

The matrix equation (\ref{KKBE}) is equivalent to a coupled system of equations for the subordinated Green functions which are derived applying Langreth's rules to the right hand side of Eq.~(\ref{KKBE}), e.g. \cite{Haug_Jauho}, and have the form (for compactness we suppress space and spin variables)
\begin{align}
\label{Dyson_diff} 
\bigl\{\partial_{\tau_1} - h_0\bigr\}\; G^M(\tau_1-\tau_2) \ &= \ \delta(\tau_1-\tau_2) \; + \; \int_{0}^{\beta} d\bar\tau \; \Sigma^M(\tau_1-\bar \tau) \, G^M(\bar \tau - \tau_2) \,,\\
\label{KKBE>} \bigl\{i\partial_{t_1} - h(t_1) \bigr\} \; G^>(t_1,t_2) \ &= \ I^>(t_1,t_2) \,,\\
\bigl\{-i\partial_{t_2} - h(t_2)\bigr\}\; G^<(t_1,t_2) \ &= \  I^<(t_1,t_2) \,,\\
\bigl\{i\partial_{t_1} - h(t_1)\bigr\}\; G^\rceil(t_1,\tau_2) \ &= \  I^\rceil(t_1,\tau_2) \,,\\
\label{KKBEl} \bigl\{-i\partial_{t_2} - h(t_2)\bigr\}\; G^\lceil(\tau_1,t_2) \ &= \  I^\lceil(\tau_1,t_2) \,,
\end{align}
Complex time arguments are indicated by $\tau$ and the system has to be supplemented by the adjoint equations. The collision integrals are given by (without loss of generality, from now on we use $t_0=0$)
\begin{align}
\label{I>} 
I^>(t_1,t_2) &= \int_{0}^{t_1} d\bar t \; \Bigl[ \Sigma^R(t_1,\bar t)\,G^>(\bar t,t_2) + \Sigma^>(t_1,\bar t) \, G^A(\bar t,t_2)\Bigr]
 \, -i\int_{0}^{\beta} d\bar\tau \; \Sigma^\rceil(t_1,\bar \tau) \, G^\lceil(\bar \tau,t_2) \,, \\
\label{I<} 
I^<(t_1,t_2) &= \int_{0}^{t_2} d\bar t \; \Bigl[ G^R(t_1,\bar t) \, \Sigma^<(\bar t,t_2) + G^>(t_1,\bar t) \, \Sigma^A(\bar t,t_2)\Bigr]
\, -i\int_{0}^{\beta} d\bar\tau \; G^\rceil(t_1,\bar \tau) \, \Sigma^\lceil(\bar \tau,t_2)\,, \\
\label{I|} 
I^\rceil(t_1,\tau_2) &= \int_{t_0}^{t_2} d\bar t  \; \Sigma^R(t_1,\bar t) \, G^\rceil(\bar t,\tau_2)
 \, +\int_{0}^{\beta} d\bar\tau \; \Sigma^\rceil(t_1,\bar \tau) \, G^M(\bar \tau-\tau_2)\,,
\end{align}
and $I^\lceil(\tau_1,t_2) = I^\rceil(t_1,\tau_2+\beta)$. The last terms in Eqs.~(\ref{I>}-\ref{I|}) account for the evolution of initial correlations.
To shorten the notation, here we have introduced the retarded (R) and advanced (A) Green functions which are defined according to 
$G^{R/A}(t_1,t_2)= \pm \Theta[\pm(t_1-t_2)][G^>(t_1,t_2) - G^<(t_1,t_2)]$  and analogously for the self-energies.

The equations of motion have to be supplemented by initial conditions. In this paper we start from the correlated equilibrium state of the electrons in the atom which is determined by the Matsubara Green function $G^M$ -- the solution of the equilibrium Dyson equation,  Eq.~(\ref{Dyson_diff}). For a convenient numerical treatment, it is transformed into the integral form \cite{Balzer} (we use $\tau = \tau_1 - \tau_2$):
\begin{align}
\label{Dyson_int} G^M(\tau) \ = \ G^0(\tau) + \int d\bar \tau d\bar {\bar \tau} \; G^0(\tau- \bar \tau) \; \tilde \Sigma[G^M](\bar \tau - \bar {\bar \tau}) \; G^M(\bar {\bar \tau}) \,.
\end{align}
$G^0$ is a reference Green function -- in our case given by the Hartree-Fock Green function -- and $\tilde \Sigma(\tau) = \Sigma^M(\tau) - \delta(\tau) \Sigma^0$, where $\Sigma^0$ is the reference self-energy evaluated with $G^0$. The boundary conditions for the Dyson equation (\ref{Dyson_int}) are given by the Kubo-Martin-Schwinger condition, $G^M(\tau) = G^M(\tau-i\beta)$.

\subsection{Self-energy approximations}\label{sec:sigma}
\begin{figure}\label{Fig_second_Born}
\centering
 \includegraphics[width=0.7\linewidth]{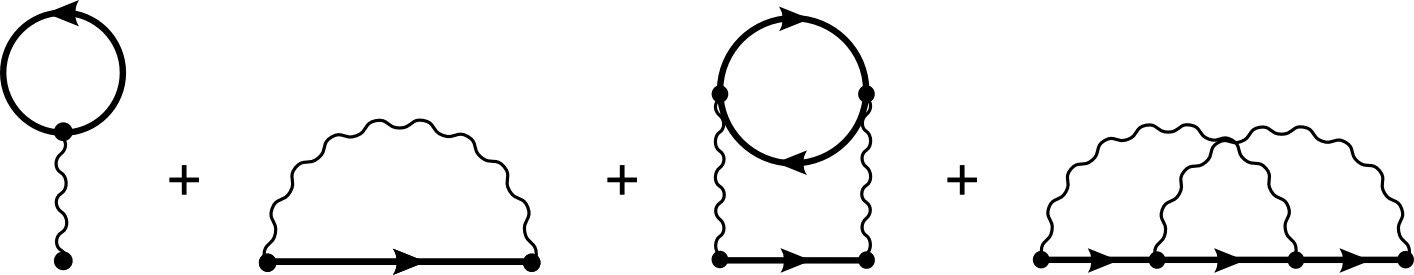}
\caption{Used approximation for the self-energy, from left to right: Hartree and Fock term, second Born and second order exchange contributions.}
\end{figure}
The equations of motion for the Green functions are formally exact if the self-energy would be known. Here we use a conserving approximation \cite{Baym_1962} for 
$\Sigma$ which includes Hartree-Fock contributions and correlations in the second Born approximation with exchange. The corresponding Feynman diagrams are shown in Fig. \ref{Fig_second_Born} and correspond to the explicit expressions (defined on the time contour)
\begin{align}\nonumber
\Sigma^{\text{HF}} (1,2) \ = \ i \, G(1,2) \, w(1,2) \; &- \; i \, \delta_{\mathcal C}(1-2) \int_{\mathcal C} d3 \, w(1,3) \, G^<(3,3) \,,\\\nonumber
\Sigma^{\text{2B}}(1,2) = i^2 \int_{\mathcal C}  d3 d4 \, G(1,3) w(1,4) G(3,4) &G(4,2) w(3,2) - i^2  \int_{\mathcal C} d3 d4 \, G(1,2) w(1,3) w(2,4) G(4,3) G(3,4) \,,
\end{align}
where we denoted $w(1,2)\equiv \delta_{\mathcal C}(1-2) w({\bf r}_1-{\bf r}_2)$. Obviously, the Hartree-Fock self-energy is time local. 

The advantage of using the method of nonequilibrium Green functions is that it provides a fully selfconsistent approach to electronic correlations in atoms in equilibrium and nonequilibrium. Solving the KKBE with the above self-energies one obtains the time evolution of a many-body system, thereby fully preserving momentum, angular momentum and total energy. Furthermore, due to the inclusion of memory effects (time integrations in the collision terms) no restriction with respect to the times apply, which is particularly important for ultrafast processes in optically excited atoms. Finally, the electromagnetic field is included non-perturbatively which allows to investigate the nonlinear dynamics of atoms in the presence of a strong excitation.

\section{Implementation}
\subsection{Basis representation}
Despite their mentioned above attractive properties, the KKBE (\ref{KKBE>}-\ref{KKBEl}) are very hard to solve numerically. Already for one-dimensional systems, they constitute a set of four-dimensional integro-differential equations (not counting spin degrees), whereas for three-dimensional systems they are even eight-dimensional. Without further approximations, this is far beyond today's numerical possibilities. A first way around this is to expand all quantities in terms of suitable single-particle basis functions $\{\phi_k\}$ with $k=1,2, \dots N_{\text{b}}$ as was demonstrated in Ref. \cite{vanleuwen_prl}. For example, the Green function and, likewise any other single-particle quantity, becomes 
an $\text{N}_{\text{b}} \times N_{\text{b}}$ dimensional matrix with the elements
\begin{align}
G_{ij}(t_1,t_2) \ = \ \int d\mathbf x_1 d\mathbf x_2 \, \phi_i^\ast(\mathbf x_1) \, G(1,2) \, \phi_j(\mathbf x_2)\,, \quad i,j=1,\dots N_{\text{b}}.
\end{align}
Below we will use for $\{\phi_k\}$ a set of $\text{N}_{\text{b}}$ orthonormal Hartree-Fock orbitals constructed from atomic orbitals. The expansion of the two-particle interaction term yields a tensor with four indices, the two-electron integrals:
\begin{align}\label{eq:w_ijkl}
w_{ijkl} = \int d\mathbf x_1 d\mathbf x_2 \, \phi_i^\ast(\mathbf x_1) \phi_j(\mathbf x_1) \, w(\mathbf x_1,\mathbf x_2) \, \phi_k^\ast(\mathbf x_2)\phi_l(\mathbf x_2) \,,
\end{align}
which are time-independent.
The main advantage of the expansion is that we have eliminated the coordinate dependencies. In terms of the basis, all equations considered earlier become equations for matrices depending on two time arguments.

So far, we have not explicitly specified the single-particle basis, the results are completely general. 
Also, the treatment of systems with different dimension is conceptually greatly simplified since it requires nothing more than a respective set of electron integrals, i.e. the matrix representation of the single-particle hamiltonian $\mathbf h_0$ and its parts $\mathbf h_{\text{pot}},\mathbf h_{\text{kin}}$, the overlap matrix $O_{ij}=\int d\mathbf x \, \phi_i^\ast (\mathbf x) \phi_j(\mathbf x)$ as well as the electron repulsion integrals $w_{ijkl}$. Furthermore, for the laser excitation the dipole matrix $\mathbf d = -e \, \mathbf r$ is needed.
Up to now, we have implemented one-dimensional numerical orbitals, three-dimensional numerical orbitals for central potentials, Slater type orbitals for atoms and Gaussian type orbitals for arbitrary molecules. Within this work we will illustrate the method for a one-dimensional model atom.

\subsection{Solution procedure}
We briefly outline of the solution procedure. We follow the techniques developed earlier, see e.g. \cite{bonitz_jpcm96,kwong_prl00,vanleuwen_prl} and references therein,  a detailed description of the algorithm is given in Refs.~\cite{Balzer} and \cite{Dahlen_Dyson}.
\begin{enumerate}
  \item A single-particle basis $\{\phi_k\}$ is chosen, and the one- and two-electron integrals (\ref{eq:w_ijkl}) are calculated. 
  \item This provides the input for a Roothaan-Hartree-Fock calculation \cite{Szabo_Ostlund} yielding the HF energies and orbitals.
  \item The electron integrals are transformed to the HF basis \cite{Hurley1988}, and the reference Green function is set up on a uniform power mesh, an adapted imaginary time-grid, as $G^0_{ij}(\tau) = \delta_{ij} \, n_i \, \mathrm e^{-(\epsilon_i-\mu)\tau}$, for $\tau \in [-\beta,0]$. $n_i$ is the occupation number of the HF-orbital $i$, which is determined by the Fermi distribution.
  \item The Dyson equation (\ref{Dyson_diff}) is solved iteratively, until a self-consistent solution for the correlated Matsubara Green function $G^M$ is found. Thus the equilibrium problem is solved.
  \item The function $G^M$ determines the initial conditions of the time propagation given by the KKBE, Eqs. (\ref{KKBE>}-\ref{KKBEl}) according to
\begin{align}
G^>(0,0) \ &= \ i\,  G^M(0^-)\,, \qquad \qquad \,  G^>(0,0) \ = \ i\, G^M(0^+) \ = \ -i\, G^M(-\beta)\,,    \\
G^\rceil(0,-i\tau) \ &= \ i\,  G^M(-\tau)\,, \qquad \qquad  G^\lceil(-i\tau,0) \ = \ i \,  G^M(\tau) \ = \ -i\,  G^M(\tau-\beta)\,.
\end{align}
  \item The KKBE are rewritten in terms of the time-evolution operator $U(t+\Delta t,t)=\mathrm e^{-ih(t) \Delta t}$ and are solved in the two-time plane by standard techniques for ordinary differential equations. We have currently implemented a fourth-order Runge-Kutta scheme \cite{Numerical_Recipes}. Due to the symmetry $G^\gtrless(t_1,t_2) = [G^\gtrless(t_2,t_1)]^\dagger$ and the boundary condition $G^>(t,t) = G^<(t,t)-i$, it is sufficient to propagate the lesser Green function in the upper triangle $t_1\geq t_2$  of the real-time plane and the greater Green function in the lower triangle $t_1<t_2$. The mixed Green functions are both propagated in the whole real-complex time-plane, though, in the long-time limit the initial correlations decay.
\end{enumerate}


\section{Simulation results}

\begin{figure}\label{fig:model}
\centering
 \includegraphics[width=0.8\linewidth]{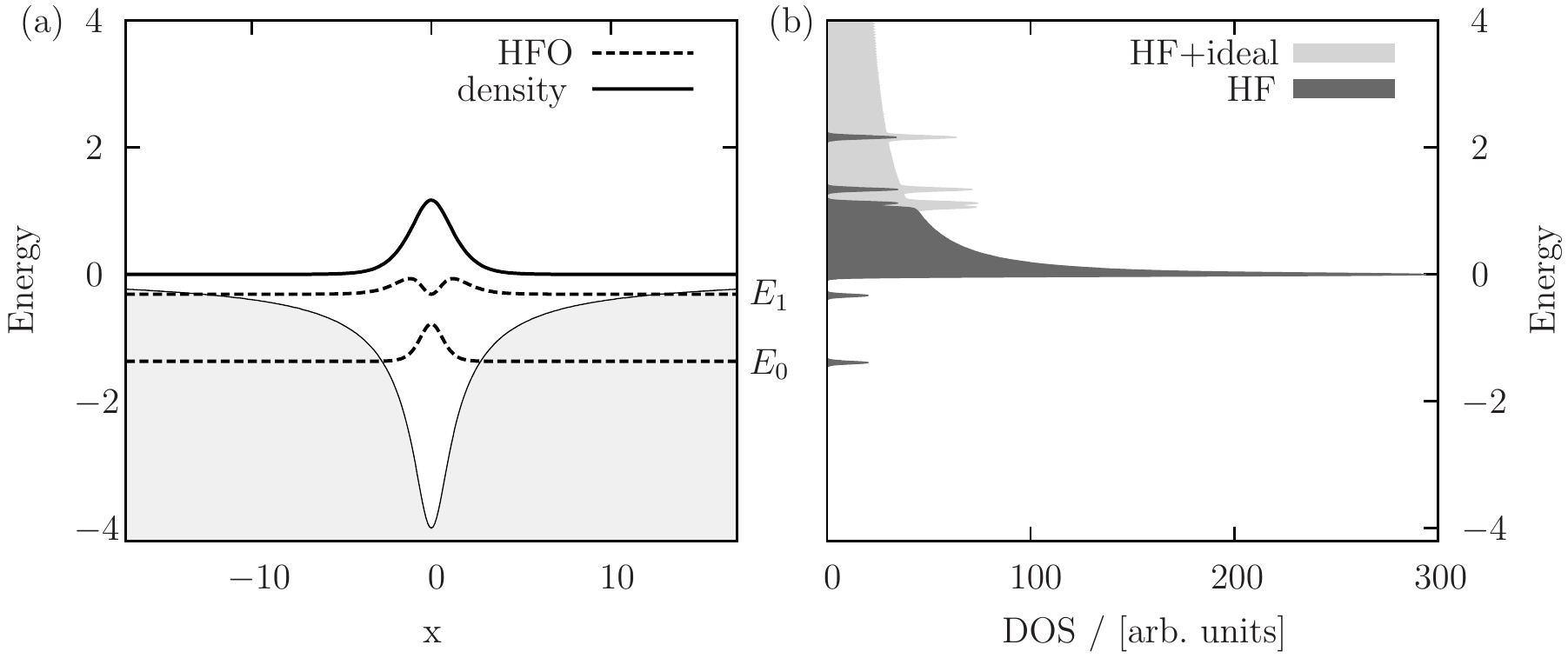}
\caption{{\bf (a)} Schematic view of the $1\rm D$ model Beryllium atom. Shown is the square of the lowest two doubly occupied Hartree-Fock orbitals (HFO) and the ground state density. {\bf (b)} Density of states (DOS) for $\text{N}_\text{b}=100$ Hartree-Fock basis functions obtained after convolution with a Gaussian of $\rm{FWHM}=0.02$. Grey curve is the result for $100$ HF basis functions and additional $200$ ideal basis function to better resolve the continuum, see Sec.~\ref{s:approx}.}
\end{figure}

\subsection{One-dimensional model atom}
To examine the presented formalism, we consider a one-dimensional model atom defined by a regularized Coulomb potential, $v(x) \ = \ - Z(x^2+\kappa_-^2)^{-1/2}$, where $Z$ is the atomic number. The screening-parameter $\kappa_{-}$ is introduced to avoid the computational difficulties arising from the singularity at the origin. Likewise, the two-particle interaction is modified according to $w(x_1,x_2)\ = \ [(x_1-x_2)^2+\kappa_+^2]^{-1/2} \ $.
This model has been used successfully in many studies of atom-laser interaction where also the influence of the choice of the screening parameters has been investigated. Here we follow Ref. \cite{Kulander1987} and use 
$\kappa_-=\kappa_+=1$.

We illustrate the method by considering beryllium ($\text{N}=\text{Z}=4$). A sketch of the confinement and the HF basis functions shifted by the orbital energies is shown in Fig. \ref{fig:model} (a). As can be seen, there are two bound states at the energies $E_0=-1.371$ and $E_1=-0.312$, each occupied by two electrons. In Fig. \ref{fig:model} (b) the density of states (DOS) is plotted, which is obtained by a modeling the continuum by a box of width $200$ a.u. and after convolution of the delta-peaks with a Gaussian of $\rm{FWHM}=0.02$. It shows the typical $1/\sqrt{E}$-decay.

\begin{table}
\begin{center}\label{tab:e0}
\begin{tabular}{ c | c c c}
\hline
\hline
\parbox[0pt][1.6em][c]{0cm}{} $\quad N_{\text{b}} \quad$ & \textbf{HF} & \textbf{2ndBorn} & \textbf{CI} \\
\hline
\parbox[0pt][1.6em][c]{0cm}{} $15$ & $-6.7390$ & $-6.7694$ & $-6.7831$ \\
\parbox[0pt][1.6em][c]{0cm}{} $30$ & $-6.7393$ & $-6.7706$ &  \\
\parbox[0pt][1.6em][c]{0cm}{} $80$ & $-6.7395$ & $-6.7710$ &  \\
\hline
\hline
\end{tabular}
\caption{Ground state energies of the $1\rm D$ Be model atom obtained from Hartree-Fock, second Born and Configuration Interaction calculations using different basis dimensions $N_b$.}
\end{center}
\end{table}

\subsection{Ground state properties}\label{ssec:groundstate}
Results for the ground state energies computed in HF and second Born approximation are presented in Table \ref{tab:e0} and compared to an exact diagonalization (CI) calculation. For the present comparison it is sufficient to consider a CI calculation using a basis of size $\text{N}_{\text{b}} = 15$ which is performed sufficiently fast. For the solution of the Dyson equation, a maximum number of $\text{N}_{\text{b}} = 80$ basis functions has been used. Together with the scaling properties, which are barely affected by the particle number but mostly by the number of basis functions, we are able to go far beyond the region accessible by sophisticated (full) CI methods \cite{Rontani2005}. To obtain the ground state within the finite temperature formalism, an inverse temperature of $\beta=100$ is used.

As can be seen, already the HF energies are close to the CI results. The inclusion of correlations on the second Born level yields a further improvement, accounting for $69 \, \% $ of the correlation energy. This confirms the trend observed for the ground states of real atoms \cite{Dahlen_Dyson}.
In this reference, it is also demonstrated how to obtain the ionization energies from the Green function using the Extended Koopmans theorem. A corresponding calculation for the one-dimensional Beryllium model yields a first ionization energy of $I_p=0.303$, which should be compared to the Hartree-Fock ionization potential from the (conventional) Koopmans theorem, $I_p=0.312$, which is known to overestimate the ionization energy.

\subsection{Time-dependent ionization dynamics following a short UV pulse}
Let us now consider the perturbation of the atom by different electromagnetic pulses. We use a Gaussian pulse (\ref{eq:gauss}) with a fixed number of cycles ($\tau=10\pi/\omega$) and an amplitude $\mathcal{E}_0=0.1$.  We consider three frequencies: $\omega_1 = |E_1|/2$, $\omega_2 = (|E_0|+|E_1|)/2$ and $\omega_3 = 1.2 \cdot |E_0|$. A classification of the different laser pulses is given by the Keldysh-parameter \cite{Keldysh1965} $\gamma = \sqrt{I_p/2 U_p}$, where $I_p$ is the ionization potential, and the ponderomotive potential is given by $U_p = \mathcal{E}_0^2/4\omega^2$. The system is propagated to $T = 200$ in Hartree-Fock approximation with $\text{N}_{\text{b}}=100$ HF basis functions.
\begin{figure}[ht]\label{fig:occupation}
\centering
 \includegraphics[width=0.9\linewidth]{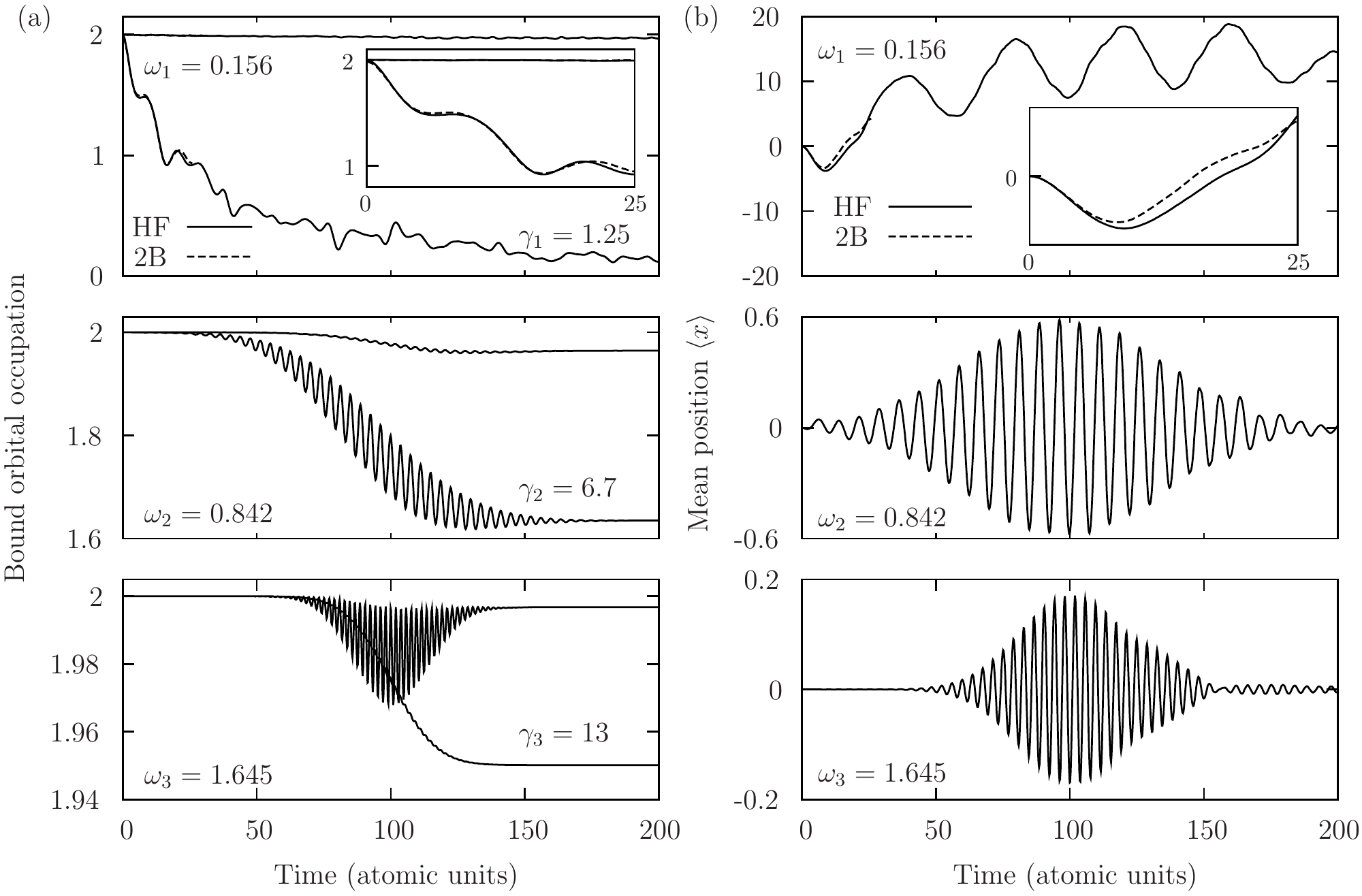}
\caption{Dynamics of the system induced by three different pulses with electric field strength $E=0.1$ and three different frequencies in Hartree-Fock (HF) and second Born (2B )approximation. For HF a basis of dimension $\text{N}_{\text{b}}=100$ was used. For 2B $\text{N}_{\text{b}}=20$ and the system was propagated up to $T=25$ a.u. {\bf (a)} Time-dependent occupation numbers of the occupied Hartree-Fock orbitals, the upper (lower) curve refers to the first (second) orbital. {\bf (b)} Time-dependent expectation value of position. The insets in the upper plots show the comparison between HF and 2B more in detail.}
\end{figure}
Fig. \ref{fig:occupation} shows the time-dependent results for the level populations during the pulse, as well as the time-dependent expectation values of the electron coordinate $\langle x \rangle$. In the upper pictures ($\omega_1 = |E_1|/2 = 0.156$) the frequency is too small to allow for PI of an electron. Nevertheless, strong ionization of the upper electrons is observed which is due to tunnel ionization ($\gamma_1=1.25$). 
The expectation value of the electron position, Fig. \ref{fig:occupation} (b), confirms this interpretation.
%
For the frequency $\omega_2 = 0.842$ PI of the upper electrons and, for the case $\omega_3 = 1.645$, direct PI of electrons from both levels is possible. In these cases, the Keldysh parameters belonging to the second orbital is $\gamma_2=6.7$ and $\gamma_3=13$, respectively, thus tunneling is not relevant, but multi-photon ionization occurs.

We have also performed various correlated simulations of the PI dynamics using the second Born approximation for the selfenergy. However, they require a propagation of the Greens functions in the full two-time plane by solving the KKBE, making the calculation computationally very costly. The upper row of Fig. \ref{fig:occupation} also contains correlated results for the dynamics up to $t=25$. During the initial phase the deviations form TDHF are still small.

\section{Approximate treatment of continuum states}\label{s:approx}
The main restriction in the time-dependent description of PI on the level of the Born approximation is the very large CPU time and memory requirement, the result shown in Fig.~\ref{fig:occupation} was obtained using $300$ time steps with $N_b=20$ basis functions. The calculations took $48$ hours on a single CPU and required 4 GB of main memory. For the first part -- the equilibrium calculations -- the basis representation saves a lot of numerical effort compared to a solution in coordinate representation. HF orbitals turn out to provide an efficient basis also for the correlated equilibrium Green function, allowing to restrict the basis to a dimension of the order $\text{N}_{\text{b}} \leq 100$ since normally only the lowest orbitals give the dominant occupation. In nonequilibrium, however, it is generally not possible to truncate the basis since, in principal, every orbital can be occupied during the excitation process. In particular, PI will lead to occupation of continuum states, and at high intensities, the electron energies may become very large \cite{bauch_pra08}.
Obviously, a fully correlated description of the whole continuum is prohibitive. At the same time, an electron ``born'' in the continuum with large kinetic energy will be only very slightly disturbed by correlations with the other electrons. This naturally suggests to develop an approximation scheme which is based on a sub-division of the basis into low and high lying orbitals which are treated with different levels of accuracy with respect to many-body effects. As a result we may hope that the ionization dynamics of the atom can be resolved sufficiently well, both in energy and in time, including electronic correlations with the required accuracy.

Our ansatz is the following: We divide the basis $\{\phi_k\}$ in three sub-systems, $\text{N}_{\text{b}}=N^{\rm corr} + N^{\rm HF}+N^{\rm id}$. The first sub-system $(1)$ contains the $\text{N}^{\rm corr}$ energetically lowest orbitals (for example the lowest atomic bound states) and is treated fully correlated. The second $(2)$ contains a number of $\text{N}^{\rm HF}$ low-lying continuum orbitals which are treated in Hartree-Fock approximation whereas the third $(3)$ of dimension $\text{N}^{\rm id}$ is treated without any particle-particle interaction. In all three sub-systems the single-particle contributions, such as the external electromagnetic field are included exactly, thus fully taking into account non-linear effects. For a given $\text{N}_{\text{b}}$, the sub-division of the basis and the numbers $\text{N}^{\rm corr}, N^{\rm HF}$ of functions in the basis parts are arbitrary and can be adapted to the considered atom and excitation conditions.

Let us now introduce this sub-division into the nonequilibrium Green functions scheme. All Green functions (all Keldysh components) and the single particle hamiltonian now become $3 \times 3$ matrices,
\begin{align}
\mathbf G \ = \ \begin{pmatrix} \mathbf G_{11} & \mathbf G_{12}  & \mathbf G_{13} \\ \mathbf G_{21} & \mathbf G_{22} & \mathbf G_{23} \\ \mathbf G_{31} & \mathbf G_{32} & \mathbf G_{33} \end{pmatrix} \ , \qquad
\mathbf h \ = \ \begin{pmatrix} \mathbf h_{11} & \mathbf h_{12} & \mathbf h_{13} \\ \mathbf h_{21} & \mathbf h_{22} & \mathbf h_{23} \\ \mathbf h_{31} & \mathbf h_{32} & \mathbf h_{33} \end{pmatrix}\,.
\end{align}
The self-energy which, in the following, is separated into a time-diagonal Hartree-Fock and a two-time correlation part, is intrinsically of the same structure. Following the idea of our approach, approximations will be introduced by systematically neglecting certain blocks of the self-energy matrix:
\begin{align}
\mathbf \Sigma^{\text{HF}} \ = \ \begin{pmatrix} \mathbf \Sigma^{\text{HF}}_{11} & \mathbf \Sigma^{\text{0}}_{12} & 0 \\ \mathbf \Sigma^{\text{0}}_{21} & \mathbf \Sigma^{\text{0}}_{22} & 0 \\ 0 & 0 & 0 \end{pmatrix} \ , \qquad
\mathbf \Sigma^{\text{corr}} \ = \ \begin{pmatrix} \mathbf \Sigma^{\text{corr}} & 0 & 0 \\ 0 & 0 & 0 \\ 0 & 0 & 0 \end{pmatrix}\,,
\end{align}
where $\Sigma^{\text{0}}$ denotes the HF selfenergy evaluated with uncorrelated Greens functions.
With this, the KKBE attain the form
\begin{align}
\label{KKBE_approx} 
\left\{ i\partial_{t_1} - \mathbf {\bar h}(t_1) \right\} \mathbf G(t_1,t_2) \ = 
\delta_{\mathcal C}(1-2) \, +
\ \left[ \int_{\mathcal C} \; dt_3 \ \begin{pmatrix} \mathbf \Sigma^{\text{corr}} \mathbf G_{11} & \mathbf \Sigma^{\text{corr}} \mathbf G_{12} &  \mathbf \Sigma^{\text{corr}} \mathbf G_{13}  \\ 0 & 0 & 0 \\ 0 & 0 & 0 \end{pmatrix}\, \right] (t_1,t_2) \, ,
\end{align}
where the HF self-energy has been included in the mean-field hamiltonian $\mathbf {\bar h}(t_1) := \mathbf h(t_1) +\mathbf \Sigma^{\text{HF}}(t_1)$ on the l.h.s. With these notations, $\mathbf G_{11}$
contains the Green functions of electrons occupying atomic bound states or undergoing transitions between low lying bound states whereas the information about ionization processes is contained in $\mathbf G_{12}$ and $\mathbf G_{13}$. Electrons in the continuum are described by $\mathbf G_{22}$ and $\mathbf G_{33}$.
This model is closely related to the Bloch equations of atomic physics or their generalizations to semiconductor optics, for a formulation using nonequilibrium Green functions, see e.g. \cite{Haug_Jauho,kwong_pss98}. 
A similar scheme has recently been reported in Refs. \cite{stefanucci} and \cite{myohanen}, where it was applied to quantum transport.

If the ionization is weak, as is normally the case with an (X)UV pulse produced from an optical laser via high harmonics, the ionization components $\mathbf G_{12}$ and $\mathbf G_{13}$ will be much smaller than contributions from occupied orbitals in $\mathbf G_{11}$. We then may expect that correlation effects in $\mathbf G_{12}$ and $\mathbf G_{13}$ play a minor role. This allows us to further simplify the model by setting $\ \mathbf \Sigma^{\text{corr}} \mathbf G_{12} \approx \ \mathbf \Sigma^{\text{corr}} \mathbf G_{13} \approx 0 \ $ on the right hand side of Eq. (\ref{KKBE_approx}). This, obviously, becomes questionable in the case of intense fields, such as optical or IR probe beams etc., but this question is beyond the scope of this work and will be considered in a forthcoming analysis.
With this approximation, only the Green function of system $(1)$ has to be considered correlated and only $\mathbf G_{11}$ has to be evolved on the full two-time plane whereas the other Green functions are completely determined by the information on the time-diagonal.

With these approximations we can write down the final system of equations to be solved as
\begin{eqnarray}
\nonumber
i\partial_{t_1}\mathbf G_{11}(t_1,t_2) - \sum_{j=1}^3 \mathbf {\bar h}_{1j}(t_1)\,\mathbf G_{j1}(t_1,t_2)
&=& 
\delta_{\mathcal C}(t_1,t_2) \, +
\ \int_{\mathcal C} \; dt_3 \ \mathbf \Sigma^{\text{corr}}(t_1,t_3) \mathbf G_{11} (t_3,t_2) \,,
\\
\nonumber
i\partial_{t_1}\mathbf G_{ik}(t_1,t_2) - \sum_{j=1}^3 \mathbf {\bar h}_{ij}(t_1)\,\mathbf G_{jk}(t_1,t_2)
&=& 
\delta_{\mathcal C}(t_1,t_2) \,,\qquad (i,k)=\{(1,2),(2,1), (2,2)\}\,,
\\
i\partial_{t_1}\mathbf G_{ik}(t_1,t_2) - \sum_{j=1}^3 \mathbf h_{ij}(t_1)\,\mathbf G_{jk}(t_1,t_2)
&=& 
\delta_{\mathcal C}(t_1,t_2) \,,\qquad i \quad\mbox{or}\quad k = 3.
\end{eqnarray}

\begin{figure}[t]\label{fig:approximation}
\centering
 \includegraphics[width=0.75\linewidth]{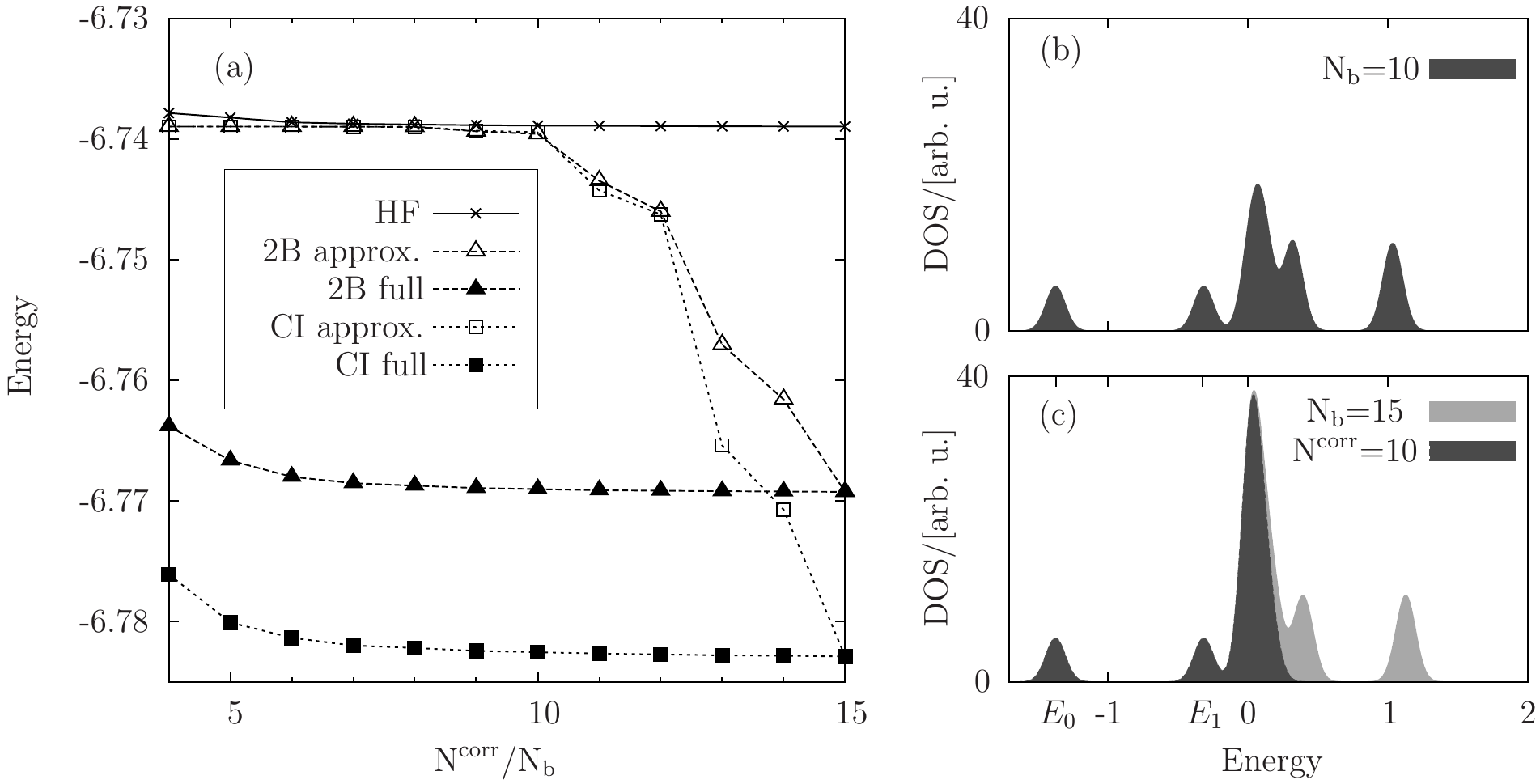}
\caption{{\bf (a)} Ground state energies obtained with the approximation scheme (open symbols), full HF (crosses), full CI (filled squares) and full second Born calculations (filled triangles) versus basis size.
For the full calculations the abscissa indicates the number of basis functions. For the approximation scheme the abscissa denotes the number $\text{N}^{\rm corr}$ of correlated orbitals, which are extracted from a Hartree-Fock calculation with $\text{N}_\text{b}=15$. {\bf (b)} DOS for a HF calculation with 10 basis functions. {\bf (c)} DOS for a HF calculation with 15 basis function (light grey) and the approximation scheme using $\text{N}^{\rm corr}=10$ and $\text{N}^{\rm HF}=5$. Dark area show the DOS corresponding to the correlated orbitals.}
\end{figure}

We give some further information on the performance of the approximation. The limiting factor in the formalism is the number of basis functions. Based on our first tests \cite{Hochstuhl_diplom} we expect that our scheme is capable to treat correlated subsystems of a size of the order of $\text{N}^{\text{corr}}=50$. This is sufficient to describe the nonequilibrium ionization dynamics of many atoms and small molecules. The size of the Hartree-Fock subsystem can be substantially larger since it depends only on a single time-argument. There the limiting factor is the computational effort of the two-electron integrals, which grows as $\mathcal O[(N^{\text{HF}})^4]$. 
Finally, the size of the ideal sub-system can, in principle, be chosen nearly as large as in standard solutions of the single-particle time-dependent Schr\"odinger equation.

Here, we present the first equilibrium results of our approximation scheme. One example is seen in Fig.~\ref{fig:model}b. where, in addition to the DOS calculated for a pure HF basis with $\text{N}^{\rm HF}=100$, we include the results for a basis supplemented by $\text{N}^{\rm id}=200$ ideal basis functions ($\text{N}^{\rm corr}=0$). The figure shows that this allows us to extend the basis to significantly higher energies and better resolve the continuum. This opens the way to investigate significantly higher excitations than before, which is particularly important for PI with (x)uv photons or for multiphoton ionization processes.
The second example is shown in Fig. \ref{fig:approximation} where we compute the ground state energy of the model atom for varying basis subdivisions, $\text{N}^{\rm corr}$ and $\text{N}^{\rm HF}$ at fixed total basis size $\text{N}^{\rm corr}+\text{N}^{\rm HF} = N_b = 15$ (for simplicity, $N^{\rm id}=0$). The results are compared with independent solutions of the Dyson equation and CI calculations for different basis sizes $N_b$ without applying the approximation scheme.
While the full calculations show fast convergence, with increasing $N_b$, the approximation scheme practically remains at the HF level for $\text{N}^{\rm corr} \leq 10$. Only for  $\text{N}^{\rm corr} > 10$ the results become better than the HF ones and eventually approach the correlated values. This behavior is not a property of the second Born approximation but is also observed in the CI calculations. 

This slow convergence is unexpected, since one would anticipate a larger HF basis to be more adequate in describing correlation effects. To analyze the reasons of this behavior we show in Fig.~\ref{fig:approximation}b) and c) the DOS for two HF basis dimensions $\text{N}^{\rm HF} =10$ and $\text{N}^{\rm HF} =15$, respectively.
As one can see, the larger basis causes an upshift of the two highest peaks of the DOS and a strong increase of the central peak around $E=0$. While this has only little effect on the HF ground state energy it strongly influences the convergence behavior of our approximation scheme. For illustration, in Fig. \ref{fig:approximation}.c) we also show the DOS for a calculation with $\text{N}^{\rm corr} =10$ and $\text{N}^{\rm HF} =5$ (dark area) which is close to the DOS for a pure HF calculation with $\text{N}^{\rm HF} =15$. Only when $\text{N}^{\rm corr}$ is so large that the correlated orbitals extend beyond the high peak is the approximation scheme approaching the correlated calculation. We are presently investigating how to avoid this unwanted behavior of our scheme in order to achieve a faster convergence. This will also be the basis for extending this approximation scheme to nonequilibrium calculations of PI processes. 


\section{Conclusion and outlook}
In this paper we applied the nonequilibrium Green functions approach to atomic photoionization. Due to the large basis size needed for the description of the continuum, which make fully correlated calculations unfeasible, we have derived an approximation scheme. Starting from the exact Keldysh-Kadanoff/Baym equations for the correlation functions we neglected certain matrix elements in the basis representation and obtained a scheme which is expected to be efficient for application to atoms. The idea is to restrict the computationally costly evaluation of the correlation self-energy to a subset of the single-particle basis. Thereby only the lowest lying bound states are treated fully correlated, while the continuum is approximated by Hartree-Fock and ideal basis functions. We have performed the first tests on the 1D-Beryllium model atom which provide a framework for further investigations. 
Further optimization of the basis subdivision towards improved convergence and application to nonequilibrium photoionizaiton dynamics will be presented in a forthcoming work.

\section*{Acknowledgements}
The authors thank R. van Leeuwen and J.W. Dufty for stimulating discussions. This work is supported by the Innovationsfond Schleswig-Holstein and by the U.S. Department of Energy award DE-FG02-07ER54946.


\end{document}